\documentclass{iaus}
\usepackage{graphicx}
\usepackage{natbib}

\title[Supernova-driven dynamo]%
{A galaxy dynamo by supernova-driven interstellar turbulence}

\author[O.~Gressel, U.~Ziegler, D.~Elstner \& G.~R\"udiger]%
{Oliver Gressel, Udo Ziegler, Detlef Elstner \& G\"unther R\"udiger}

\affiliation{Astrophysikalisches Institut Potsdam,  %
  An der Sternwarte 16, 14482 Potsdam, Germany %
  \break email:~\texttt{\{ogressel,uziegler,elstner,gruediger\}@aip.de}}

\pubyear{2009}
\volume{259}
\pagerange{100--100}
\date{Dec. 10, 2008 and in revised form ??}
\jname{Cosmic Magnetic Fields: From Planets,to Stars and Galaxies} 
\editors{K.G. Strassmeier, A.G. Kosovichev \& J.E. Beckman, eds.}

\newcommand{\rms}[1]{\left<\right.\!#1\!\left.\right>}
\newcommand{\Myr}{\,\rm Myr}

\newcommand{\kpc}{\,\rm kpc}
\newcommand{\kms}{\,\rm km\,s^{-1}}
\newcommand{\vis}{\,\rm cm^{2}s^{-1}}

\newcommand{\EMF}{\mathcal E}
\newcommand{\Rm}{\rm Rm}
\newcommand{\Pm}{\rm Pm}

\begin{document}

\maketitle

\begin{abstract}
  Supernovae are the dominant energy source for driving turbulence
  within the interstellar plasma. Until recently, their effects on
  magnetic field amplification in disk galaxies remained a matter of
  speculation. By means of self-consistent simulations of
  supernova-driven turbulence, we find an exponential amplification of
  the mean magnetic field on timescales of a few hundred million
  years. The robustness of the observed fast dynamo is checked at
  different magnetic Reynolds numbers, and we find sustained dynamo
  action at moderate $\Rm$. This indicates that the mechanism might
  indeed be of relevance for the real ISM.

  Sensing the flow via passive tracer fields, we infer that SNe
  produce a turbulent $\alpha$~effect which is consistent with the
  predictions of quasilinear theory. To lay a foundation for global
  mean-field models, we aim to explore the scaling of the dynamo
  tensors with respect to the key parameters of our simulations. Here
  we give a first account on the variation with the supernova rate.

\keywords{turbulence -- ISM: supernova remnants, dynamics, magnetic fields}
\end{abstract}


We here present new results on our local box simulations of a
differentially rotating, vertically stratified, turbulent interstellar
medium threaded by weak magnetic fields
\citep{2008AN....329..619G}. In our model, we apply optically thin
radiative cooling and heating to account for the heterogeneous,
multi-phase nature of the ISM. Improving over existing models, we
compute a radiatively stable initial solution to avoid the transient
collapse seen in models applying an isothermal stratification. The
central feature of our simulations is the driving of turbulence via
several thousand localised injections of thermal energy, which well
resemble the kinetics of the supernova feedback. Unlike for artificial
forcing, the energy and distribution of the SNe are determined by
observable parameters.

In a preceding paper, we have shown that the turbulence created by SNe
does in fact exponentially amplify the mean magnetic field
\citep{2008A&A...486L..35G}. Here we extend this work towards a
broader parameter base. The organisation of this article is as
follows: in Section~\ref{sec:slow_fast}, we discuss the possible
relevance of the discovered effect for realistic Reynolds numbers and
compare the kinetically driven SN dynamo with the cosmic ray dynamo
found by \citet{2004ApJ...605L..33H}. In Section~\ref{sec:sigma}, we
then report on the influence of the supernova rate on the measured
dynamo parameters. For a short review on mean-field modelling of the
galactic dynamo, we refer the reader to Elstner et al. (this volume).

\section{Slow versus fast dynamo} \label{sec:slow_fast}

In laminar dynamos, diffusion sets the relevant timescale for magnetic
reconnection, thus defining an upper limit for the allowable growth
rate of the mean magnetic field. Because the microscopic diffusivity
is usually low, these dynamos are commonly referred to as ``slow
dynamos''. Within the ISM, the diffusion time $\tau_{\rm d}=L^2/\eta$
(related to the microscopic value $\eta\simeq10^8\vis$) by far exceeds
the Hubble time. This means that the field amplification mechanism in
galaxies needs to be a ``fast dynamo'' in the sense that it works on a
timescale different than $\tau_{\rm d}$ \citep{1999ApJ...517..700L}.
Accordingly, one assumes that the galactic dynamo is determined by
some sort of effective turbulent diffusivity $\eta_{\rm t}$. Although
the dynamo may still be limited by topological changes via
reconnection, this process can be considerably faster due to the much
higher value of $\eta_{\rm t}$ compared to $\eta$. Turbulent
reconnection has recently been studied numerically by
Otmianowska-Mazur, Kowal \& Lazarian (this volume).

According to the definition of $\tau_{\rm d}$, in the laminar case one
expects the dynamo growth rate to increase with $\eta$. The picture,
however, changes significantly as soon as the Reynolds number is high
enough to allow for developed turbulence. Structures of integral
length scale $L$ are now efficiently broken down to the Kolmogorov
microscale where the atomic diffusion takes over. Varying the
microscopic value of $\eta$ does only change the extent of the
inertial range towards higher wavenumbers but is no longer relevant to
the large-scale flow. This implies that a turbulent dynamo should be
insensitive to variations in $\eta$ as soon as a critical value
$\Rm_{\rm c}$ is exceeded \citep[cf.][]{1999ApJ...517..700L}.

\begin{figure} 
  \center\includegraphics[width=0.95\columnwidth]{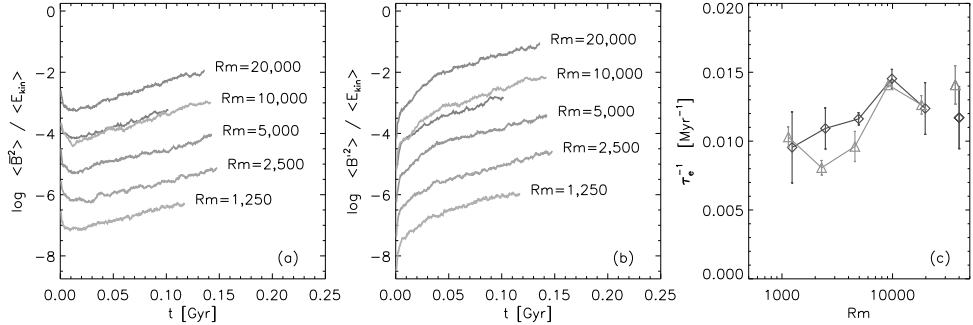}
  \caption{Evolution of the regular (a) and fluctuating (b) magnetic field
    strength for different $\Rm$. For clarity, the ordinates of the different
    models have been offset by an order of magnitude each. In panel (c), we
    compare the growth rates for the turbulent (diamonds) and regular
    (triangles) magnetic field. The unconnected data points to the right
    correspond to a run with $\eta=0$ and provide an indication for the level
    of numerical diffusivity.}
  \label{fig:fig1}
\end{figure}

In Fig.~\ref{fig:fig1}, we plot the evolution of the mean and
turbulent magnetic field (normalised to the kinetic energy) for
different magnetic Reynolds numbers $\Rm = L^2\,q\Omega/\eta$. The
values for $\Rm$ are obtained by varying $\eta$ while keeping the
rotation rate $\Omega$ fixed. The kinematic viscosity $\nu$ is
furthermore adopted to keep the magnetic Prandtl number
$\Pm=\nu/\eta=2.5$ constant. Irrespective of the value of $\Rm$, we
observe a nice and steady exponential amplification of both the
regular and turbulent fields. To estimate the influence of the finite
grid resolution, we have performed a fiducial run (dark grey lines in
panels (a) and (b) of Fig.~\ref{fig:fig1}) at double the grid spacing
for $\Rm=10,000$. The obtained values are consistent with the higher
resolved run and provide a first indication that the simulation
results are reasonably converged at this level of dissipation and
below.

In panel (c) of Fig.~\ref{fig:fig1}, we compare the growth rates for
the turbulent (diamonds) and regular (triangles) magnetic field as a
function of the magnetic Reynolds number. The unconnected data points
to the right correspond to a run with $\eta=0$, i.e., the case of
(formally) infinite $\Rm$. As can be seen from a comparison with these
points, above $\Rm\simeq10,000$ we are limited by the finite value of
the inherent numerical diffusivity of our code, i.e., better resolved
runs become mandatory to study the regime of higher $\Rm$. With
respect to the reliably converged runs, we observe growth rates that
increase with $\Rm$ -- suggesting that the effect is of genuinely
turbulent nature.

In their models of the cosmic-ray-driven buoyant instability,
\citet{2007ApJ...668..110O} observe that their dynamo crucially relies
on the presence of a ``microscopic'' diffusivity $\eta$. The authors
actually find the efficiency of the field amplification to scale with
this parameter. Moreover, they state that no scale separation is
manifest in the spectra of their simulations. In this respect, it
remains disputable whether the high value for $\eta$ might rather be
interpreted as an effective turbulent dissipation $\eta_{\rm t}$. If
so, the simulations would have to be regarded as large eddy
simulations, i.e., simulations assuming turbulent effects rather than
having them emerge from first principles.

\section{The dependence on supernova rate} \label{sec:sigma} 

Quasilinear theory \citep{1980mfmd.book.....K} is a powerful tool in
predicting the dynamo $\alpha$~effect from the underlying kinetic
structure of the turbulent flow. Unfortunately, as far as galaxies are
concerned, little has been known with certainty about the vertical
profiles of the turbulent velocity $u'(z)$ and the mean flow
$\bar{u}_z(z)$. In particular, $\bar{u}_z$ had to be ignored in
analytical derivations \citep[see e.g.][]{1996A&A...311..451F}
assuming hydrostatic equilibrium. In that regard, direct simulations
can give important new insights, especially if one is interested in
the dependence on factors like the supernova frequency.

\begin{figure}   
  \begin{minipage}[t]{0.68\columnwidth}
    \includegraphics[width=0.91\columnwidth,bb=0 185 316 350,clip]{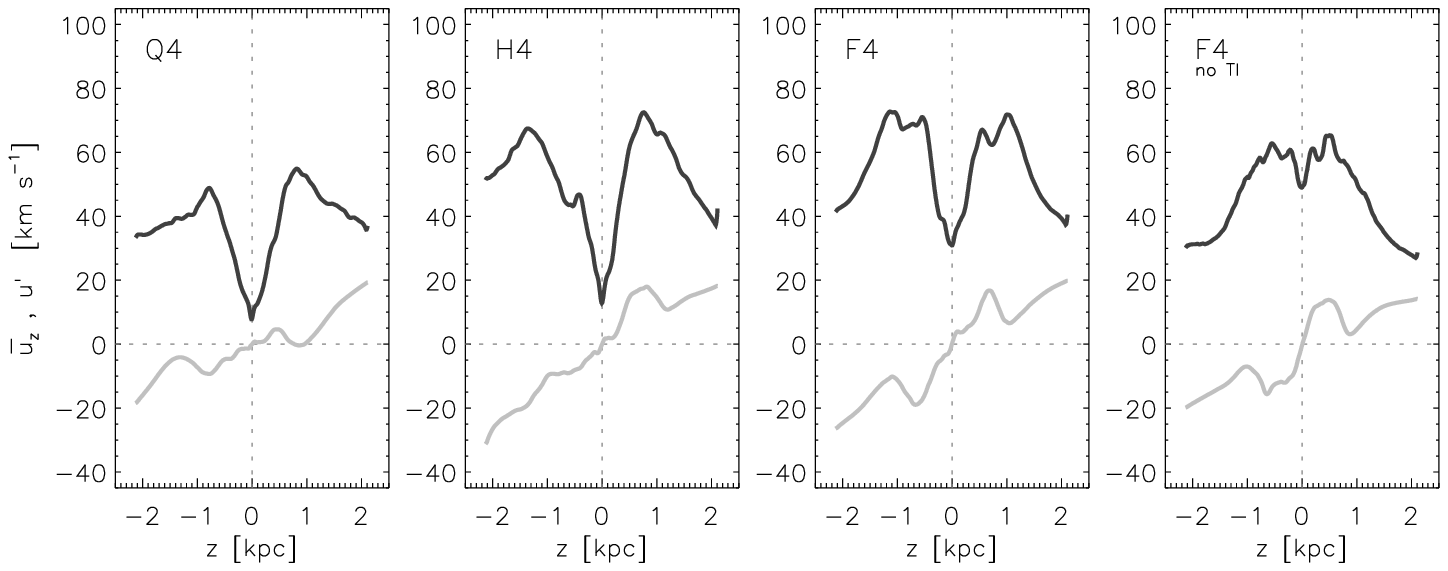}
  \end{minipage}
  \begin{minipage}[b]{0.305\columnwidth}
    \caption{Vertical profiles of the mean flow $\bar{u}_z$ (light) and the
      turbulent velocity $u'$ (dark colour) at quarter, half, and full
      supernova rate $\sigma/\sigma_0$.\vspace{10pt}}
    \label{fig:fig2}
  \end{minipage}
\end{figure}

\subsection{Vertical structure \& wind} 

In the following, we measure the supernova rate $\sigma$ in units of
the galactic value $\sigma_0=30\Myr^{-1}\kpc^{-2}$, representative of
type-II SNe. All models include sheared galactic rotation with
$|q\Omega|= 100\kms\kpc^{-1}$. The time averaged turbulent and mean
velocity profiles of three models with $\sigma/\sigma_0=0.25$, $0.5$,
and $1.0$ are depicted in Fig.~\ref{fig:fig2}: The vertical structure
of the turbulent velocity $u'$ shows a distinct M-shape, which peaks
at $\pm1\kpc$. The positive gradient of $u'$ in the central disk
strongly suggests an inward transport of the mean magnetic field. The
inner part of the profiles is similarly shaped as the ones obtained
from MRI turbulence \citep{2004A&A...423L..29D,2007ApJ...663..183P}
but considerably steeper. Crudely extrapolating the fall-off in
$u'(z)$, we estimate that the MRI might become important in
maintaining the observed velocity dispersions above galactic heights
of $|z| \simeq 3\kpc$.

\begin{figure} 
  \begin{minipage}[t]{0.68\columnwidth}
    \includegraphics[width=0.82\columnwidth]{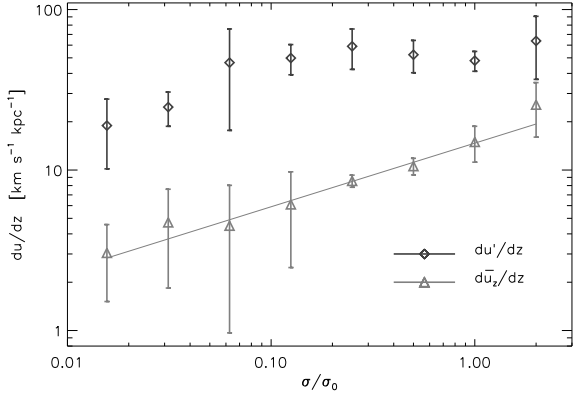}
  \end{minipage}
  \begin{minipage}[b]{0.31\columnwidth}
    \caption{Vertical velocity gradients as a function of the
      supernova rate $\sigma/\sigma_0$. The values are derived from
      the inner disk for $u'$, and from the full domain for
      $\bar{u}_z$ (cf. Fig.~\ref{fig:fig2}); the overplotted
      regression shows a logarithmic slope of $0.4$.\vspace{10pt}}
    \label{fig:fig3}
  \end{minipage}
\end{figure}

While the overall amplitude of the turbulence increases with the SN
intensity, its gradient is only weakly affected. This can be seen in
Fig.~\ref{fig:fig3}, where we plot the scaling of the fitted
velocity gradients. Unlike the turbulent velocity gradient ${\rm
d}u'/{\rm d}z$, which goes into saturation for $\sigma\simeq
0.1\,\sigma_0$, the wind profile shows a distinct scaling with the
supernova frequency. We thus estimate the wind profile from kinetic
feedback as
\begin{equation}
  \bar{u}_z(z) \quad\simeq\quad
  15.\kms\,\left(\frac{\sigma}{\sigma_0}\right)^{0.4}\,\frac{z}{1.\kpc}\,,
\end{equation}
which, of course, neglects the characteristic modulation of the mean
flow within the V-shaped region of $u'(z)$, where the kinetic pressure
counteracts the thermal pressure.

\subsection{Dynamo tensors \& growth rates} 

\begin{figure} 
  \center\includegraphics[width=\columnwidth]{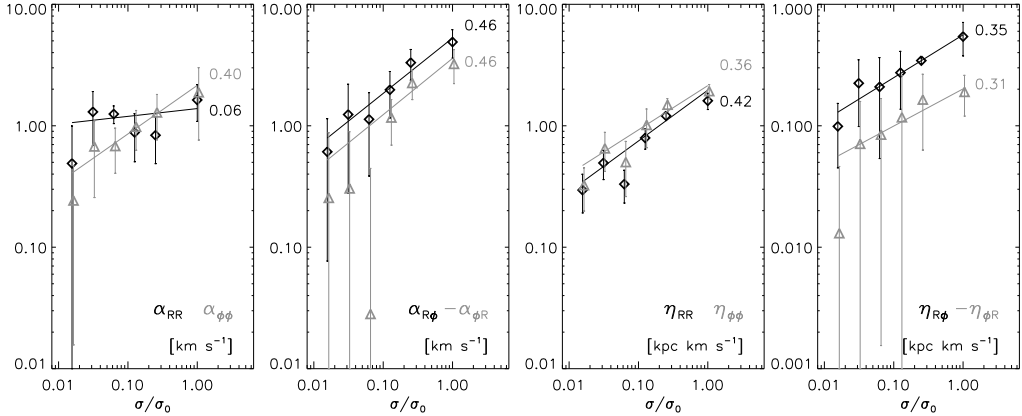}
  \caption{Coefficients of the dynamo $\alpha$ and $\tilde{\eta}$ tensor as
    obtained with the test-field method.}
  \label{fig:fig4}
\end{figure}

One main focus of our work is the derivation of mean-field closure
parameters from direct simulations. We here assume a standard
parameterisation of the turbulent EMF
\begin{equation}
  \EMF_i = \alpha_{ij} \bar{B}_j 
         - \tilde{\eta}_{ij}\varepsilon_{jkl}\partial_k \bar{B}_l\,,
  \quad i,j \in \left\{R,\phi\right\}, k=z\,,
  \label{eq:param}
\end{equation}
which enters the mean-field induction equation via $\nabla\times
\EMF$. To measure these coefficients, we apply the test-field method
of \citet{2005AN....326..245S,2007GApFD.101...81S}. From the thus
derived profiles $\alpha(z)$ and $\tilde{\eta}(z)$, we compute
vertically averaged amplitudes. In Fig.~\ref{fig:fig4}, we plot these
integral mean values as a function of the supernova rate
$\sigma$. Notably, for each of the data points we had to perform a
separate 3D simulation, covering a few hundred million years to obtain
reasonable statistics. Despite the strong scatter in the plotted
values, we observe a quite robust scaling with respect to $\sigma$. An
exception to this are the coefficients $\alpha_{RR}$, which display
irregular scaling in some of the runs and will be subject to further
investigations.

The main result from Fig.~\ref{fig:fig4} is that all coefficients
(except $\alpha_{RR}$) scale in a similar way. In particular, this
implies that the dynamo number
\begin{equation}
  C_{\alpha}=\alpha_{\phi\phi} H/\eta_{\rm t}\,,
\end{equation}
remains approximately constant with $\sigma$. Furthermore, as we have
already demonstrated in an earlier work, the effects of the
off-diagonal elements $\alpha_{R\phi}$ and $\alpha_{\phi R}$ (which
are responsible for the diamagnetic pumping) are approximately
balanced by the mean flow $\bar{u}_z$. Comparing the slope shown in
Fig.~\ref{fig:fig3} with the ones in the second panel of
Fig~\ref{fig:fig4}, we see that this balance of forces is roughly
independent of $\sigma$ and therefore seems to be of rather general
nature.

\begin{figure} 
  \begin{minipage}[t]{0.67\columnwidth}
    \includegraphics[width=0.82\columnwidth]{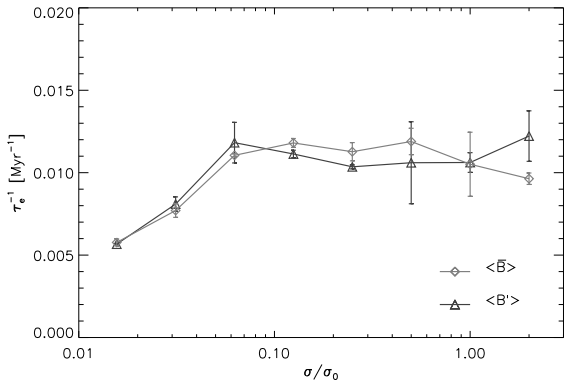}
  \end{minipage}
  \begin{minipage}[b]{0.32\columnwidth}
    \caption{Growth rate $\tau^{\!-1}_{\rm e}$ of the (vertically integrated)
      regular field $\rms{\bar{B}}$ (diamonds) and the turbulent field
      $\rms{B'}$ (triangles) as a function of $\sigma/\sigma_0$.\vspace{10pt}}
    \label{fig:fig5}
  \end{minipage}
\end{figure}

With the constant ratio of the $\alpha$~effect versus the turbulent
diffusion $\eta_{\rm t}$ on one hand, and the mutual balance of the
turbulent pumping and mean flow on the other hand, we expect the
overall growth rates to be largely insensitive to the applied
supernova rate. For sufficiently high values of $\sigma$ this is
indeed the case, as can be seen from Fig.~\ref{fig:fig5} above -- we
want to point out that these growth rates are obtained from the direct
simulations and hence do not depend on the $\alpha$~prescription.
With constant growth rates over more than one order of magnitude in
star formation activity, the kinetically driven SN dynamo proves to be
rather universal. Accordingly, it may well be applicable to a wide
range of scenarios -- covering dwarf galaxies as well as starbursts.

\section{Conclusions}

We have presented recent simulation results on the variation of the
key dynamo parameters with the supernova frequency. Yet more work is
needed to understand how the vertical structure of the galactic disk
is linked to the emergence of the dynamo effect and the associated
transport processes. As a primary question, one is of course
interested in finding a simple explanation for the observed scaling
relations -- which are somewhat flatter than the square-root of the
supernova rate $\sigma$.

From quasilinear theory, it is understood that the dynamo effect
depends on vertical gradients both in the density $\rho$ and the
turbulence intensity $u'$. The disk structure, on the other hand, is
determined by the balance of the kinetic pressure from the forcing and
the gravitation of the stellar component -- here equipartition
arguments can yield a first approximation. To connect kinematic
properties with dynamo parameters, we however need to specify the
correlation time $\tau_{\rm c}$ of the turbulent flow. By comparing
direct simulations with SOCA predictions, it will eventually become
possible to determine how this quantity depends on the supernova
frequency.

To connect the measured dynamo parameters $\alpha$ and $\tilde{\eta}$
with the growth rates observed in the simulations -- and thereby check
the applicability of the approach -- more mean-field modelling is
necessary. We are, however, confident that the classical framework of
MF-MHD remains a valuable tool in the quest of understanding galactic
magnetism.

\begin{acknowledgments}
This project was kindly supported by the Deutsche
Forschungsgemeinschaft (DFG) under grant Zi-717/2-2. All computations were
performed at the AIP \texttt{babel} cluster.
\end{acknowledgments}

\def\aap{\textit{A\&A} }
\def\apj{\textit{ApJ} }
\def\an{\textit{AN} }
\def\gapfd{\textit{GApFD} }

\begin{discussion}

  \discuss{Johansen}{}
  \discuss{Gressel}{}

  \discuss{Brandenburg}{}
  \discuss{Gressel}{}

  \discuss{de Gouveia Dal Pino}{}
  \discuss{Gressel}{}

\end{discussion}

\end{document}